\documentstyle[12pt,aaspp4,epsf]{article}

\begin{document}

\title{Chaos and the Shapes of Elliptical Galaxies}

\author{D. Merritt}
\affil{Department of Physics and Astronomy, Rutgers University,
    New Brunswick, NJ 08855}
\bigskip
\centerline{Rutgers Astrophysics Preprint Series No. 186}
\bigskip
\centerline{To appear in {\it SCIENCE}, Jan. 19, 1996}
\bigskip
\begin{abstract}
{\it Hubble Space Telescope} (HST) observations reveal that the 
density of stars in most elliptical galaxies rises toward the 
center in a power-law cusp.
Many of these galaxies also contain central dark objects, 
possibly supermassive black holes.
The gravitational force from a steep cusp or black hole
will destroy most of the box orbits that constitute the 
``backbone'' of a triaxial stellar system.
Detailed modelling demonstrates that the resulting chaos can 
preclude a self-consistent, strongly triaxial equilibrium.
Most elliptical galaxies may therefore be nearly axisymmetric, 
either oblate or prolate.
\end{abstract}


Information about the three-dimensional shape of a galaxy is lost 
when the galaxy is projected onto the plane of the sky.
This loss of information is acute in the case of elliptical galaxies, 
whose apparent shapes are elliptical but whose intrinsic shapes 
could be oblate, prolate or fully triaxial.

Before about 1975, elliptical galaxies were thought to be rotationally 
flattened oblate spheroids.
The discovery that elliptical galaxies rotate much more slowly 
than a fluid body with the same shape (1) led to the hypothesis that 
most of these systems are triaxial ellipsoids, with shapes that 
are maintained by anisotropic velocity dispersions rather than by 
centrifugal force (2).
The triaxial hypothesis was supported by the successful construction of 
self-consistent triaxial models on the computer (3).
Most of the stars in these numerical models occupied ``regular'' orbits that 
respect three isolating integrals, two in addition to the energy; 
the major families of regular orbits are the short- and long-axis ``tubes,'' 
and the ``boxes'' (4). 
Box orbits are uniquely associated with the triaxial geometry; 
they densely fill a box- or bowtie-shaped region, and a star on such an 
orbit passes arbitrarily close to the galaxy center after many 
oscillations.
The time-averaged shape of a box orbit mimics that of the 
underlying galaxy, and the potential from a star on a box orbit
helps support the triaxial shape of the galaxy as a whole.
Box orbits are always found to be strongly populated 
in the self-consistent triaxial models.

Box orbits exist only in triaxial potentials with ``cores,'' that is,
models in 
which the density near the center is approximately 
constant and the corresponding gravitational potential is roughly 
quadratic in the coordinates (5).
Recent HST observations of the centers of elliptical galaxies (6) 
reveal that these galaxies almost never have constant-density 
cores; the stellar density always continues to rise, 
roughly as a power law, into the smallest observable radius.
In fainter ellipticals, the stellar density increases roughly as 
$\rho\propto r^{-2}$, while for brighter ellipticals the cusp 
slope is $\rho\propto r^{-1}$ or shallower (7).
In addition, there is increasingly strong evidence for massive dark 
objects (MDO's), possibly supermassive black holes, at the 
centers of many elliptical galaxies (8).
In the most convincing cases, these central singularities appear 
to contain as much as 1\% of the total mass of the galaxy. 

A central mass concentration can strongly perturb the motion of a 
star on a box orbit, regardless of its apocenter distance, 
since a star on such an orbit will eventually pass arbitrarily close to the 
center and be deflected by the strong gravitational force there 
(9).
The result is a sensitive dependence of the orbital trajectory on 
initial conditions; in other words, the orbit loses its two 
non-classical integrals of motion and becomes chaotic.
The degree of chaos can be quantified via the Liapunov 
characteristic numbers that measure the average rate of exponential 
divergence of two initially nearby trajectories.
Fig. 1a shows histograms of Liapunov numbers for ensembles of 
box-like orbits (defined as orbits that have a stationary point)
at one energy in a triaxial model with the density law
$$
\rho(m) = \rho_0(m^2+m_0^2)^{-1}(1+m^2)^{-1}, \ \ \ \ 
m^2 = {x^2\over a^2} + {y^2\over b^2} + {z^2\over c^2}, \eqno (1)
$$
with $c/a=0.5$ and $b/a=0.79$.
The parameter $m_0$ is a ``core radius''; when $m_0=1$, Eq. 1 
reduces to the ``Perfect Ellipsoid'' in which all orbits are 
regular (10), while for small $m_0$, this density law has an 
$r^{-2}$ central cusp similar to those observed in many elliptical galaxies.

When $m_0< 10^{-1}$, one finds that most of the box-like orbits 
are chaotic; the only exceptions are orbits which lie close to stable, 
resonant orbits that avoid the center (Fig. 2a).
The destruction of the box orbits also occurs in models with a 
central singularity, or ``black hole'' (Fig. 1b, 2b).
In both cases, the typical Liapunov time scale for divergence of 
nearby trajectories is 3-5 times the 
oscillation period of the long-axis closed orbit of the same 
energy, henceforth defined as the ``dynamical time'' (Table 1).
Stars in the central regions of elliptical 
galaxies have made $10^2 - 10^3$ radial oscillations since the 
epoch of galaxy formation; thus, elliptical galaxies are many
Liapunov times old.

A star on a chaotic orbit will eventually visit every point 
in configuration space consistent with energy conservation;
it will fill a region inside of the equipotential surface 
corresponding to its energy.
These surfaces are more nearly spherical than the equi{\it 
density} 
surfaces that define the galaxy figure; hence, chaotic 
orbits are less useful than regular box orbits for building 
self-consistent models.
However, even chaotic orbits have structure.
Fig. 3 illustrates the density of an ensemble of 
stars that fills chaotic phase space in an approximately time-independent 
way at one energy.
The shape is similar to that of a superposition of box-like 
orbits, since the chaotic trajectory fills that part of phase 
space that would have been occupied by box orbits in a fully 
integrable model.
But because the chaotic phase space at a given energy is 
interconnected via
the ``Arnold web'' (11), there exists just a single invariant 
density like the one shown in Fig. 3 at each energy.

The ``mixing time'' of a chaotic orbit may be defined as 
the time required for an ensemble of stars on that orbit to reach 
a fully-mixed state like that of Fig. 3.
In a real galaxy, the mixing process is likely to be extremely complex, 
involving violent collapse and rapidly varying forces during galaxy 
formation (12).
However we can place an upper limit on the mixing time by asking 
how long it takes an ensemble of stars in the chaotic phase space 
of a time-{\it in}dependent potential to fill its allowed 
phase-space region in a nearly uniform way.
Numerical experiments show that this relaxation process 
is roughly exponential, with a time constant of
about 100 dynamical times in triaxial potentials with steep cusps or 
massive central singularities (13).
After a few hundred dynamical times, the density in the chaotic 
phase-space region achieves a nearly constant, coarse-grained 
value and ceases to evolve.
We would therefore expect the chaotic orbits in at least the 
central regions of a triaxial galaxy with a strong 
central mass concentration to be fully mixed.

The invariant distribution of Fig. 3 plays the role of a single orbit: 
it represents an unchanging and irreducible distribution of stars and can 
be used as a building block for a self-consistent 
model.
However -- unlike the regular orbits, which respect three 
integrals of the motion and therefore comprise a 
two-parameter family at every energy -- there is only one 
invariant density at each energy in chaotic phase space.
The replacement of the regular box orbits by chaotic trajectories 
thus limits the freedom to construct a self-consistent 
model, since it effectively reduces the number of different orbits.
This limitation does not exist in oblate or prolate 
geometries, however, since axisymmetric potentials support only
tube orbits, all of which avoid the center and most of which remain 
regular.
Thus the nonexistence of a triaxial equilibrium with a given 
density profile would imply that a galaxy with the same 
mass distribution must be either axisymmetric, or in the process 
of evolving toward an axisymmetric state.

The degree to which chaos limits the freedom to construct 
triaxial equilibria was explored via two models with Dehnen's (14)
density law:
$$
\rho(m)=\rho_0 m^{-\gamma}(1+m)^{-(4-\gamma)}. \eqno (2)
$$
The first model explored (the ``strong cusp'' model) had $\gamma\approx 2$, 
corresponding to fainter elliptical galaxies like M32.
The ``weak cusp'' model had $\gamma\approx 1$, a good description 
of brighter ellipticals like M87.
The axis ratios were chosen to be $c/a=0.5$ and $b/a=0.79$.
A total of 7000 orbits were integrated for 100 dynamical times in 
each of the models, and their time-averaged densities stored in a 
grid of $10^3$ cells.
A quadratic-programming algorithm was then used to find a set of 
non-negative orbital weights that reproduced the known mass of 
the model in the cells (15).

Attempts to construct self-consistent solutions using just the 
regular orbits failed for both mass models.
Quasi-equilibrium solutions -- in which chaotic orbits, computed for 
just 100 orbital periods, were included with arbitrary orbital 
weights -- were found to exist for both weak- and strong-cusp 
models.
However, real galaxies constructed in this way would 
evolve near the center as the chaotic orbits mixed toward 
their invariant distributions at each energy (16).

More nearly stationary solutions were successfully constructed for 
the weak-cusp mass model; all of the chaotic orbits within the 
inner half-mass radius could be replaced by the smaller set of 
invariant distributions without violating self-consistency.
However these models could not be made fully mixed at both 
large and small radii.
No significant fraction of the mass could be placed on 
fully-mixed chaotic orbits in the strong-cusp model without 
driving the solution away from self-consistency.
The greater freedom to find solutions in the weak-cusp case resulted 
from the larger number of regular orbit families in this 
potential, which allowed less weight to be placed on the 
chaotic orbits.

These attempts to find self-consistent equilbria were based on 
only two, strongly triaxial mass models; more nearly axisymmetric 
models with the same density profile would presumably be easier to construct.
However these results demonstrate that chaos can severely 
reduce the size of solution space for triaxial models and, at 
least in some cases, preclude self-consistent equilibria.
It is reasonable to conclude that strongly triaxial configurations 
are difficult to construct in a self-consistent way, and that 
real ellipticals are either axisymmetric or nearly axisymmetric.

Although no attempts have yet been made to construct 
self-consistent triaxial models with central ``black 
holes,'' the results shown here in Figs. 1 and 2 suggest that chaos 
would constrain such solutions about as strongly as it constrains
triaxial models with steep density cusps.
Secure detections of MDO's have been made in 
at least two elliptical galaxies (17,18); the kinematical 
signature in both cases was a high streaming velocity of stars or 
gas very near the center.
In addition, strong kinematical evidence for MDO's has been found
in a number of S0 galaxies and spirals with bulges (8).
The failure to detect MDO's in some ellipticals may be due 
to the lack of a rotating subpopulation, or to the fact that 
the galaxy is not oriented in such a way that the rotation is 
easily observed.
The average mass of the MDO's in these three 
galaxies is $\sim 0.0045$ when expressed in units of the total 
stellar mass of the galaxy.
This is close to the average mass required per galaxy if MDO's
are dead quasars (19), which suggests that most or all 
ellipticals may harbor nuclear black holes.

If most elliptical galaxies contain steep density cusps, nuclear 
black holes, or both, then the arguments given above suggest that 
axisymmetry would generally be preferred over triaxiality for
these galaxies.
This would be especially true for lower-luminosity ellipticals
which have the steepest cusps and the shortest dynamical times
on average (6).
The axisymmetric hypothesis is difficult to confirm since no unambiguous test 
for triaxiality exists.
However, most recent studies of elliptical galaxy intrinsic 
shapes have found that few if any elliptical galaxies need to be 
strongly triaxial (20).

The dependence of observed rotation rate on flattening in low-luminosity 
ellipticals has long been known to be consistent with 
oblate symmetry for these galaxies (21).
A classical test for triaxiality is dependence of major-axis 
orientation on radius (22).
Such ``isophote twists'' are seen in a number of elliptical 
galaxies, but their interpretation is complicated by the 
likelihood that some of the twisted galaxies are not relaxed, 
while in others the twist may result from misaligned disks and 
bars.
A stronger test for triaxiality is 
the detection of stellar streaming along the apparent 
minor axis of a galaxy (22).
Minor axis rotation is rare, however, and a statistical study of the 38 
elliptical galaxies for which two-dimensional velocity 
data are available suggests that the data can be well fit by a 
distribution in which 60\% of galaxies are oblate and 40\% 
prolate (23).

The shapes of a few elliptical galaxies have been constrained by 
detailed comparison of numerical models with kinematical data.
The best example is M32, the dwarf companion to the 
Andromeda galaxy.
M32 has a $\rho\propto r^{-1.6}$ stellar cusp and also shows 
convincing evidence for a MDO containing $\sim 0.25\%$ of the 
total galaxy mass (17); 
thus one would expect axisymmetry to be strongly preferred for 
this galaxy.
In fact, oblate models reproduce the detailed kinematics of M32
extremely well (24).
Rings and disks of gas or dust can sometimes be used as tracers of the 
shape of the gravitational potential in elliptical galaxies (25), although 
few of these subsystems are both extended and regular enough that 
the results are convincing.
However the kinematics of a neutral 
hydrogen ring surrounding the elliptical galaxy IC 2006 suggests 
that its dark halo is accurately axisymmetric (26).

Bright ellipticals are observed to be slowly rotating, and if these
galaxies are generically axisymmetric, their slow rotation implies
nearly equal numbers of stars on tube orbits travelling in both 
directions.
Such a configuration would arise naturally as the potential evolved 
from triaxiality into axial symmetry via the mechanism described 
here: stars on box-like orbits undergo 
periodic changes in the direction of their angular momenta, and 
an ensemble of such stars would presumably end up populating a 
set of tube orbits with roughly equal numbers of rotating and 
counter-rotating members.
A galaxy with this orbital composition might reveal itself 
via a strongly flattened or double-peaked distribution of 
line-of-sight velocities (27). 

\bigskip

\vfill\eject

\begin{table}
\caption{Liapunov numbers of boxlike orbits in triaxial potentials.
Orbits were computed for $10^4$ dynamical times 
at the half-mass energy in the potential corresponding to Eq. 1, 
with $c/a=0.5$ and $b/a=0.79$.
Liapunov numbers $\sigma_1$ and $\sigma_2$ are given in units of 
the dynamical time; $M_{BH}$ represents black hole mass in units of 
the total mass of the model.}

\begin{tabular}{rlcc}
$m_0$ & $M_{BH}$ & $\sigma_1$ & $\sigma_2$ \\ \hline
$10^{-1}$ & -- & $0.14\pm 0.06$    & $0.045\pm 0.02$ \\
$10^{-2}$ & -- & $0.21\pm 0.09$    & $0.078\pm 0.03$ \\
$10^{-3}$ & -- & $0.27\pm 0.05$    & $0.085\pm 0.02$ \\
$10^{-1}$ &      $10^{-3}$         & $0.15 \pm 0.03$ & $0.066\pm 0.02$ \\
$10^{-1}$ &      $3\times 10^{-3}$ & $0.20 \pm 0.04$ & $0.097\pm 0.02$ \\
$10^{-1}$ &      $10^{-2}$         & $0.28 \pm 0.08$ & $0.16 \pm 0.04$ \\ \hline

\end{tabular}
\end{table}

\vfill\eject
\centerline{REFERENCES}
\bigskip

1. F. Bertola and M. Capaccioli, {\it Astrophys. J.} {\bf 200}, 439 
(1975).

2. J. J. Binney, {\it Mon. Not. R. Astron. Soc.} {\bf 183}, 501 
(1978).

3. M. Schwarzschild, {\it Astrophys. J.} {\bf 232}, 236 (1979).

4. An illustration of the major orbit families in integrable 
triaxial models may be found in D. Merritt, {\it Science} {\bf 
259}, 1867 (1993).

5. J. Lees and M. Schwarzschild, {\it Astrophys. J.} {\bf 384}, 
491 (1992).

6.  Kormendy, J. {\it et al.}, in {\it ESO/OHP Workshop on Dwarf Galaxies},
G. Meylan and P. Prugniel, Eds. (European Southern Observatory, Garching,
1995), p. 147.

7. D. Merritt and T. Fridman, in ASP Conf. Ser. Vol. 50, 
{\it Fresh Views of Elliptical Galaxies}, A. Buzzoni, 
A. Renzini and A. Serrano, Eds. (Astronomical Society of the Pacific,
San Francisco, 1995), p. 13.

8. J. Kormendy and D. Richstone, {\it Annual Review of Astronomy 
and Astrophysics}, {\bf 33} (1995).

9. O. E. Gerhard and J. J. Binney, {\it Mon. Not. R. Astron. Soc.} 
{\bf 216}, 467 (1985).

10. G. G. Kuzmin, in {\it Dynamics of Galaxies and Clusters}, T. B. 
Omarov, Ed. (Akademie Nauk Kazakh SSR, Alma Ata, 1973), p. 71;
P. T. de Zeeuw and D. Lynden-Bell, {\it Mon. Not. R. Astron. Soc.} 
{\bf 215}, 713, (1985).

11. A. J. Lichtenberg and M. A. Lieberman, {\it Regular and 
Stochastic Motion} (Springer-Verlag, New York, 1989).

12. D. Lynden-Bell, {\it Mon. Not. R. Astron. Soc.} {\bf 136}, 
101 (1967).

13. D. Merritt and M. Valluri, in preparation.

14. W. Dehnen, {\it Mon. Not. R. Astron. Soc.} {\bf 265}, 250 
(1993).

15. D. Merritt and T. Fridman, {\it Astrophys. J.} {\bf 456} 
(1996).

16. M. Schwarzschild, {\it Astrophys. J.} {\bf 409}, 563 (1993).

17. A. Dressler and D. Richstone, {\it Astrophys. J.} {\bf 324}, 
701 (1988); van der Marel, R. P., de Zeeuw, T., Rix, H.-W., 
White, S. D. M., {\it Mon. Not. R. Astron. Soc.} {\bf 271}, 99.

18. R. J. Harms {\it et al.}, {\it Astrophys. J.} {\bf 435}, L35 
(1994).

19. A. Soltan, {\it Mon. Not. R. Astron. Soc.} {\bf 200}, 115 
(1982).

20. D. Merritt, in {\it 
Morphological and Physical Classification of Galaxies}, G. Longo, 
M. Capaccioli, G. Busarello, Eds. (Kluwer Academic, Norwell, MA, 
1992), pp. 309-320.

21. R. L. Davies, G. Efstathiou, S. M. Fall, G. Illingworth, and P. 
L. Schechter, {\it Astrophys. J.} {\bf 266}, 41 
(1983).

22. G. Contopoulos, {\it Zeitschrift f\"ur Astrophysik} {\bf 39}, 
126 (1956).

23. M. Franx, G. D. Illingworth and P. T. de Zeeuw, {\it 
Astrophys. J.} {\bf 383}, 112 (1991).

24. E. E. Qian, P. T. de Zeeuw, and C. Hunter, 
{\it Mon. Not. R. Astron. Soc.} {\bf 274}, 602 (1995).

25. F. Bertola {\it et al.}, {\it Astrophys. J.} {\bf 373}, 369 
(1991).

26. M. Franx and T. de Zeeuw, {\it Astrophys. J. Letters} {\bf 
392}, L47 (1992).

27. O. Gerhard, {\it Mon. Not. R. Astron. Soc.} {\bf 265}, 213 
(1993).

28. M. Valluri carried out the calculations on which Fig. 3 was 
based, and made helpful comments on the manuscript.
S. Tremaine, as referee, also made a number of suggestions that
improved the presentation.
This work was supported by NSF grant AST 93-18617 and by NASA grant NAG 
5-2803.

\clearpage

\figcaption[]
{Histograms of Liapunov numbers for iso-energetic 
ensembles of box-like orbits in triaxial potentials.
Starting points for the orbits were chosen from a uniform grid on 
an equipotential surface near the half-mass radius of the model (Fig. 2).
Each orbit was integrated for $10^4$ dynamical times $T_d$; the 
thick curve represents the largest Liapunov number $\sigma_1$, 
the thin curve the second Liapunov number $\sigma_2$.
Regular orbits lie in the narrow peaks near $\sigma T_d=0$.
(a) $m_0=10^{-3}$; (b) $m_0=10^{-1}$, and a central point mass 
containing 0.3\% of the total galaxy mass has been added.}

\figcaption[]
{Starting points of the regular and chaotic orbits
whose Liapunov numbers make up the histograms of Fig. 1a and 1b.  
Each dot represents an initial point on one octant of 
the equipotential surface; small
dots are chaotic orbits and large dots are regular orbits.
Every orbit was dropped with zero velocity from this surface.
The $X$ [$Z$] axes are the long [short] axes of the figure.}

\figcaption[]
{Fig. 3. Invariant density of an isoenergetic ensemble of 
5000 stars in the chaotic phase space of a triaxial model
with $m_0=10^{-3}, c/a=0.5$ and $b/a=0.79$.
The $X$ [$Z$] axis is the long [short] axis of the triaxial figure.
Plotted are the densities near each of the three principal 
planes.}

\plotone{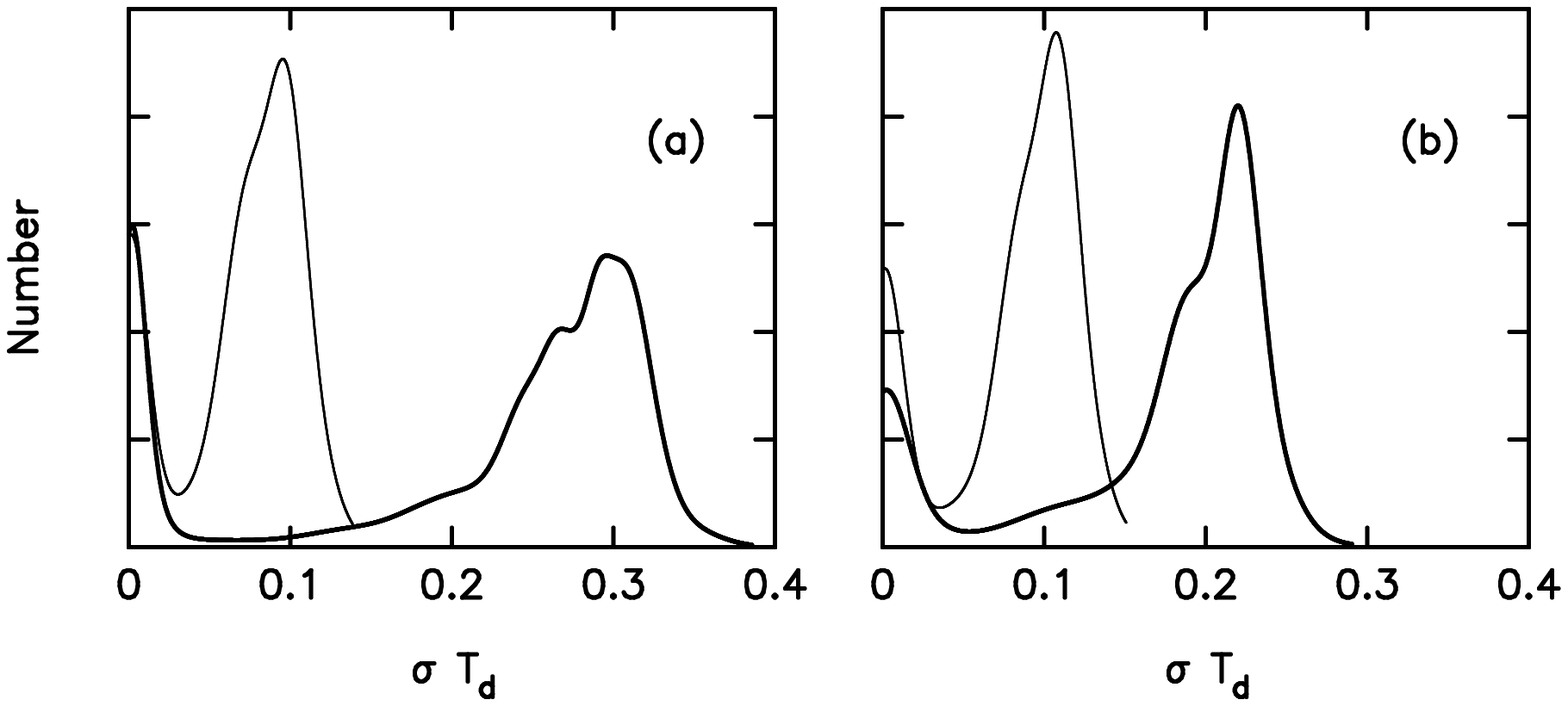}

\plotone{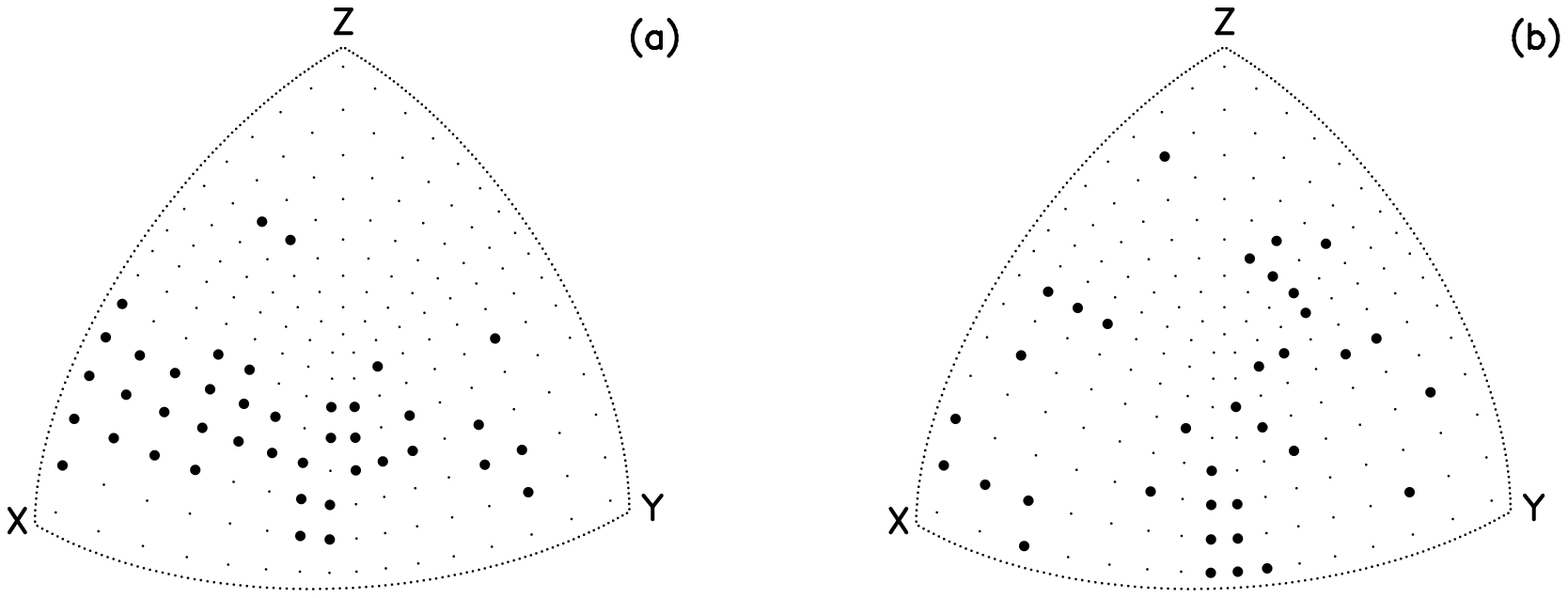}

\plotone{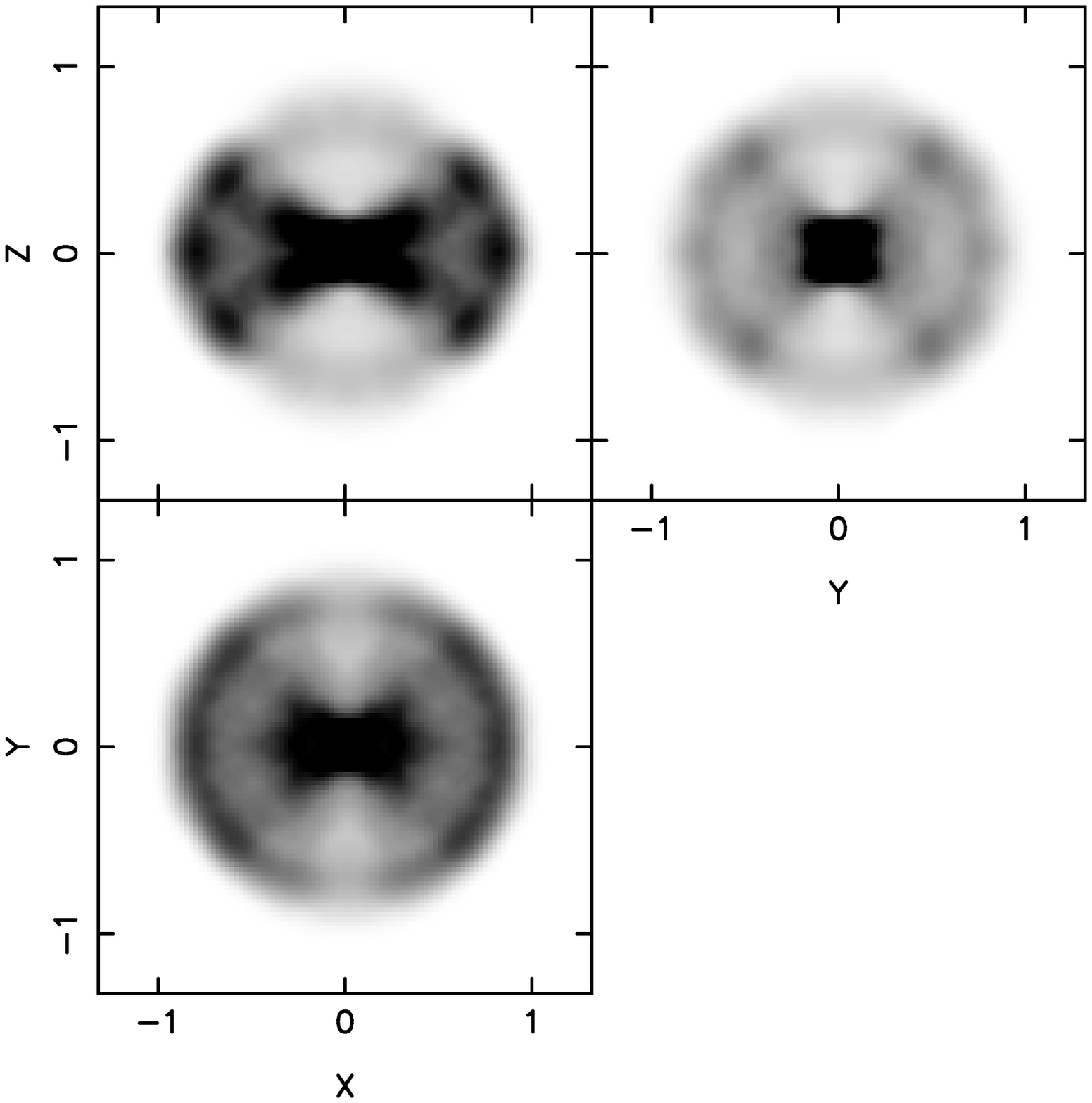}
 
\end{document}